\documentclass[
   ,final            
  ,numberedheadings 
 , article                
  ]
  {aipproc}

\layoutstyle{6x9}

\usepackage{equation}

\begin{document}

\title{Construction of Lie Superalgebras from
 Triple Product Systems}

\author{Susumu Okubo}{
  address={Department of Physics and Astronomy, University of Rochester,
  Rochester, NY 14627}
}



\begin{abstract}
Any simple Lie superalgebras over the complex field can be
constructed from some triple systems. Examples of Lie
superalgebras $D(2,1;\alpha)$, $G(3)$ and $F(4)$ are given by
utilizing a general construction method based upon $(-1,-1)$
balanced Freudenthal-Kantor triple system.
\end{abstract}

\maketitle


\section{Lie and Anti-Lie Triple Systems}

The triple products are perhaps a little unfamiliar in physics,
although it has been utilized to find some solutions of
Yang-Baxter equation \cite{okubo1} as well as of
para-statistics~\cite{okubo2}. Some other examples are also found
in reference 3.

Before going into details, let us briefly sketch what a triple
product is. Let $V$ be a vector space over a field $F$. Then, a
bilinear product in $V$ is a linear map:

$$V \otimes V \rightarrow V$$

\noindent denoted as $xy = z$ for $x$, $y$, \ $z \ \in\ V$. If
$e_1, e_2, \dots, e_N$ is a basis of $V$, then its algebraic
structure is completely determined by its multiplication table of

\begin{equation}
e_j e_k = \sum^N_{\ell =1} C^\ell_{jk} e_\ell \qquad (j,k, \ell =
1,2, \dots, N)\label{eq:oneone}
\end{equation}

\noindent for some structure constants $C^\ell_{jk} \ \in \ F$.

In contrast, a triple product defined in $V$ is a linear mapping

$$V \otimes V \otimes V \rightarrow V$$

\noindent and we write the triple product as $xyz$, $[x,y,z]$ or
$x \cdot y \cdot z$ or any other symbol you would prefer. Then,
the analogue of Eq. ({\ref{eq:oneone}}) is

\begin{equation}
\left[ e_j , e_k , e_\ell \right] = \sum^N_{m=1} C^m_{jk \ell} e_m
\label{eq:onetwo}
 \end{equation}

\noindent for some structure constants $C^m_{jk \ell} \ \in\ F$,
where we used the symbol of $[x,y,z]$ as the triple product here
to be definite.

A simple example \cite{kamiya} of a triple system is obtained as
follows. Let $< .|.>$ be a bilinear form in $V$, satisfying a
condition of

\begin{equation}
<y|x>\ = - \epsilon <x|y>, \quad (x,\ y\ \in\ V)
\label{eq:onethree}
\end{equation}

\noindent $\epsilon = \pm 1$. We now introduce a triple product in
$V$ by

\begin{equation}
[x,y,z]:\  = \ <x|z>y \ + \epsilon <y|z>x. \label{eq:onefour}
\end{equation}

\noindent It is easy to verify that it satisfies

\begin{subequations}\label{foo}
\begin{eqnarray}
&{\rm (i)}& \quad [x,y,z] = \epsilon [y,x,z] \label{foo-a}
\label{eq:onefivea}\\
\noalign{\vskip 6pt}%
 &{\rm (ii)}& \quad [x,y,z] + [y,z,x] + [z,x,y] = 0 \label{foo-b} \label{eq:onefiveb}\\
 \noalign{\vskip 6pt}%
&{\rm (iii)}& \quad [u,v,[x,y,z]] = [[u,v,x],y,z] + [x,[u,v,y],z]
+ [x,y,[u,v,z]] \label{foo-c} \label{eq:onefivec}
\end{eqnarray}
\end{subequations}

\noindent for any $u,\ v,\ x,\ y,\ z\ \in\ V$. We then say that
any vector space $V$ with a triple product $[x,y,z]$ satisfying
Eqs. (1.5) is a Lie \cite{lister} (for $\epsilon = -1$) and an
anti-Lie \cite{faulkner} (for $\epsilon = +1$) triple system,
respectively.

As we will see shortly, Lie and anti-Lie triple systems are
intimately related to Lie and Lie superalgebras, respectively.

\section{Canonical Construction}

\setcounter{equation}{0}

It is well-known \cite{kamiya}-\cite{kak} that we can construct
Lie and Lie superalgebras, respectively from Lie and anti-Lie
triple systems as follows.

We first introduce the Lie-multiplication operator $L(.,.): V
\otimes V \rightarrow \ {\rm End}\ V$ by

\begin{equation}
L(x,y)z:\  = [x,y,z]. \label{eq:twoone}
\end{equation}

\noindent We emphasize the fact that $L(x,y)\ \in\ {\rm End}\ V$
is a linear transformation operator in the vector space $V$, so
that they form an {\it associative} algebra in the ordinary sense.
We note then that Eq. (\ref{eq:onefivea}) immediately gives

\begin{equation}
L(y,x) = \epsilon L(x,y)\label{eq:twotwo}
\end{equation}

\noindent since for any $z\ \in\ V$, we calculate

$$\left\{ L(y,x) - \epsilon L(x,y) \right\} z = [y,x,z] - \epsilon
[x,y,z] = 0$$

\noindent by Eq. (\ref{eq:onefivea}). Secondly, Eq.
(\ref{eq:onefivec}) is rewritten as

\begin{equation}
[L(u,v),L(x,y)] = L([u,v,x],y) + L(x,[u,v,y]) \label{eq:twothree}
\end{equation}

\noindent where we have set

\begin{equation}
[L(u,v),L(x,y)]:\  = L(u,v)L(x,y) - L(x,y)L(u,v)
\label{eq:twofour}
\end{equation}

\noindent as the usual commutator. To see the validity of Eq.
(\ref{eq:twothree}) we calculate

\begin{eqnarray}
L(u,v)L(x,y)z &=& L(u,v)[x,y,z] = [u,v,[x,y,z]], \nonumber\\
\noalign{\vskip 4pt}%
L(x,y)L(u,v)z &=& L(x,y)[u,v,z] = [x,y,[u,v,z]], \nonumber\\
\noalign{\vskip 4pt}%
L([u,v,x],y)z &=& [[u,v,x],y,z], \nonumber\\
\noalign{\vskip 4pt}%
L(x,[u,v,y]) z &=& [x,[u,v,y],z],\nonumber
\end{eqnarray}

\noindent from the definition of $L(x,y)$ acting on any $z \ \in\
V$. Therefore, Eq. (\ref{eq:onefivec}) is rewritten as

\begin{equation}
\left\{ [L(u,v),L(x,y)] -L([u,v,x],y) -L(x,[u,v,y])\right\} z = 0
\label{eq:twofive}
\end{equation}

\noindent which leads to the validity of Eq. (\ref{eq:twofour})
since the linear transformation acting on $z \ \in \ V$ in the
left side of Eq. (\ref{eq:twofive}) is a null transformation. We
then note that Eq. (\ref{eq:twofour}) gives a Lie algebra since
$L(u,v)$ and $L(x,y)$ form an associative algebra.

So far, we did {\it not} utilize Eq. (\ref{eq:onefiveb}). We
consider a larger vector space

\begin{equation}
W = L(V,V) \oplus V:\  = V_{\overline 0} \oplus V_{\overline 1}
\label{eq:twosix}
\end{equation}

\noindent where $L(V,V)$ is a vector space consisting of all
linear combination of $L(x,y)$'s $(x,\ y,\ \in\ V)$. Note that the
commutators such as $[x,y]$ and $[x,L(y,z)]$ are {\it not} defined
in the theory. We can, nevertheless, introduce these commutators
formally by relations,

\begin{subequations}\label{foo}
\begin{eqnarray}
&{\rm (i)}& \quad [x,y]:\  = L(x,y) = \epsilon
L(y,x),\label{foo-a}\label{eq:twosevena}\\
\noalign{\vskip 6pt}%
&{\rm (ii)}& \quad [L(x,y),z]:\  = - [z,L(x,y)]:\  = [x,
y,z].\label{foo-b}\label{eq:twosevenb}
\end{eqnarray}
\end{subequations}

We have now to consider two cases of $\epsilon =1$ and $-1$,
separately. First, we discuss the Lie triple system with $\epsilon
=-1$. In that case, Eqs. (\ref{eq:twotwo}) and
(\ref{eq:twosevena}) give

\begin{equation}
[x,y] = -[y,x]. \label{eq:twoeight}
\end{equation}

\noindent Then, $W$ becomes a larger Lie algebra, i.e., we can
prove to have

\begin{subequations}\label{foo}
\begin{eqnarray}
& & [X,Y] = -[Y,X] \label{foo-a} \label{eq:twoninea}\\
\noalign{\vskip 6pt}%
& & \left[ [X,Y], Z\right] + \left[ [Y,Z],X\right] + \left[
[Z,X],Y\right] = 0 \label{foo-b} \label{eq:twonineb}
\end{eqnarray}
\end{subequations}

\noindent for any $X,\ Y,\ Z\ \in \ W$, if we take Eq.
(\ref{eq:onefiveb}) into account.

On the contrary, the case of the anti-Lie triple system with
$\epsilon = +1$ leads to a Lie superalgebra as follows. First,
this leads to

\begin{equation}
[x,y] = [y,x] \label{eq:twoten}
\end{equation}

\noindent instead of Eq. (\ref{eq:twoeight}). Moreover, we note
that

\begin{equation}
V_{\overline 0} = L(V,V), \quad L_{\overline 1} = V
\label{eq:twoeleven}
\end{equation}

\noindent offers a $Z_2$-graded space since

\begin{eqnarray}
& & \left[ V_{\overline 0}, V_{\overline 0} \right] \subseteq
V_{\overline 0}, \quad \left[ V_{\overline 0} , V_{\overline 1}
\right] \subseteq V_{\overline 1} , \nonumber\\
\noalign{\vskip 4pt}%
& & \left[ V_{\overline 1} , V_{\overline 1} \right] \subseteq
V_{\overline 0} \label{eq:twotwelve}
\end{eqnarray}

\noindent by Eqs. (\ref{eq:twothree}) and (2.7). Then, we can
introduce the grading function by

\begin{equation}
(-1)^X = \left\{
\begin{array}{lllll}
1, &&{\rm if} &&X = L(x,y)\ \in\ V_{\overline 0} \\
\noalign{\vskip 6pt}%
 -1, &&{\rm if} &&X = x\ \in\ V = V_{\overline 1}
 \end{array}\right. .\label{eq:twothirteen}
 \end{equation}

\noindent In this case, the resulting algebra is a Lie
superalgebra satisfying \cite{kac,sch}

\begin{subequations}\label{foo}
\begin{eqnarray}
&{\rm (i)}& \quad [X,Y] = - (-1)^{XY} [Y,X]\label{foo-a}
\label{eq:twofourteena}\\
\noalign{\vskip 6pt}%
&{\rm (ii)}& \quad (-1)^{XZ} \left[ [X,Y], Z \right] + (-1)^{YX}
\left[ [Y,Z], X\right] + (-1)^{ZY} \left[ [Z,X],Y\right] = 0
\label{foo-b} \label{eq:twofourteenb}
\end{eqnarray}
\end{subequations}

\noindent instead of Eqs. (2.9).

It may be instructive to inspect the example of Eq.
(\ref{eq:onefour}) for Lie and Lie supertriple systems. Consider
first the case of $\epsilon = -1$. Let $e_1, e_2, \dots, e_N$ $(N
= \dim V)$ be a basis of $V$ with

\begin{equation}
<e_j |e_k>\ = \delta_{jk}, \quad (j,k = 1,2, \dots , N).
\label{eq:twofifteen}
\end{equation}

\noindent Then, setting

\begin{equation}
J_{jk} = -J_{kj} = L \left( e_j , e_k \right) ,
\label{eq:twosixteen}
\end{equation}

\noindent Eqs. (\ref{eq:twoone}) and (\ref{eq:onefour}) give

\begin{equation}
J_{jk} e_\ell = \delta_{j \ell} e_k - \delta_{k \ell} e_j
\label{eq:twoseventeen}
\end{equation}

\noindent and Eq. (\ref{eq:twothree}) leads to the $so(N)$ Lie
algebra of

\begin{equation}
\left[ J_{jk}, J_{\ell m}\right] = \delta_{j \ell} J_{km} -
\delta_{k \ell} J_{jm} + \delta_{jm} J_{\ell k} - \delta_{k m}
J_{\ell j} \label{eq:twoeighteen}
\end{equation}

\noindent since we calculate

\begin{eqnarray}
\left[ J_{jk} , J_{\ell m} \right] &=& \left[ L \left( e_j , e_k
\right), L \left( e_\ell , e_m \right) \right]\nonumber\\
\noalign{\vskip 4pt}%
&=& L\left( \left[ e_j, e_k, e_\ell \right], e_m \right) + L
\left( e_\ell,  \left[ e_j , e_k , e_m \right] \right) \nonumber\\
\noalign{\vskip 4pt}%
&=& L\left( \delta_{j \ell} e_k - \delta_{k \ell} e_j, e_m \right)
+ L \left( e_\ell , \delta_{jm} e_k - \delta_{km} e_j
\right)\nonumber\\
\noalign{\vskip 4pt}%
&=& \delta_{j \ell} L \left( e_k, e_m \right) - \delta_{k \ell}
L\left( e_j , e_m \right) + \delta_{jm} L\left( e_\ell , e_k
\right) - \delta_{km} L\left( e_\ell , e_j \right).\nonumber
\end{eqnarray}

\noindent On the other side, Eqs. (2.7) give

\begin{subequations}\label{foo}
\begin{eqnarray}
& & \left[ e_j , e_k \right] = J_{j k}, \label{foo-a}
\label{eq:twonineteena}\\
\noalign{\vskip 4pt}%
& & \left[ J_{jk} , e_\ell \right] = - \left[ e_\ell, J_{jk}
\right] = \delta_{j \ell} e_k - \delta_{k \ell} e_j .
\label{foo-b} \label{eq:twonineteenb}
\end{eqnarray}
\end{subequations}

\noindent Therefore introducing

\begin{eqnarray}
J_{0j} &=& -J_{j0} = e_j , \nonumber\\
\noalign{\vskip 4pt}%
J_{00} &=& 0 , \label{eq:twotwenty}
\end{eqnarray}

\noindent for $j = 1, 2, \dots, N$, the relations of Eqs.
(\ref{eq:twoeighteen}) and (2.19) are combined into a single
relation of

\begin{subequations}\label{foo}
\begin{eqnarray}
& & \left[ J_{AB}, J_{CD} \right] = \delta_{AC} J_{BD} -
\delta_{BC} J_{AD} + \delta_{BD} J_{AC} - \delta_{AD}
J_{AC},\label{foo-a} \label{eq:twotwentyonea}\\
\noalign{\vskip 4pt}%
& & J_{AB} = - J_{BA} \label{foo-b} \label{eq:twotwentyoneb}
\end{eqnarray}
\end{subequations}

\noindent for $A,\ B,\ C,\ D = 0, 1,2,\dots , N$. Thus, the larger
Lie algebra $W$ is now $so(N+1)$.

On the other case of $\epsilon = +1$, the condition Eq.
(\ref{eq:onethree}) must be modified as

\begin{equation}
<e_j | e_k>\ = \ \in_{jk}\  = - \in_{kj} \label{eq:twotwentytwo}
\end{equation}

\noindent for $j, k = 1,2,\dots ,N$, where $\in_{jk}$ is the
symplectic form ($N =$ even). In that case, $J_{jk}$ given by Eq.
(\ref{eq:twosixteen}) now satisfies

\begin{subequations}\label{foo}
\begin{eqnarray}
& & \left[ J_{jk}, J_{\ell m} \right] =\  \in_{j \ell} J_{km}\  -
\in_{k \ell} J_{jm} \ + \in_{jm} J_{\ell k} \ - \in_{km} J_{\ell
j} \label{foo-a} \label{eq:twotwentythreea}\\
\noalign{\vskip 6pt}%
\noalign{\hbox{with}}
\noalign{\vskip 6pt}%
& & J_{jk} = J_{kj}, \label{foo-b}\label{eq:twotwentythreeb}
\end{eqnarray}
\end{subequations}

\noindent which is the symplectic Lie algebra $sp(N)$. The larger
vector space $W$ now gives the Lie superalgebra \cite{kac,sch}
$osp(1,N)$, although we will not go into detail.

In this connection, we may note that if $V$ is a super-space from
the beginning satisfying

\begin{equation}
<x|y>\ = (-1)^{xy} <y|x> \label{eq:twotwentyfour}
\end{equation}

\noindent instead of Eq. (\ref{eq:onethree}), we could have
obtained a more general Lie superalgebra $osp(M,N)$. For details,
see reference 2.

\section{Freudenthal-Kantor Triple Systems}

\setcounter{equation}{0}

It is known \cite{kamiya2,asano} that all simple Lie algebras over
the complex field can be constructed from some suitable triple
systems. Recently, we have shown \cite{kamiya3,kam} that all
simple Lie superalgebras over the complex field can  also be
constructed from triple systems. To this end, we must consider
more general triple systems. As an example, we briefly sketch the
notion of $(\epsilon, \epsilon)$ balanced Freudenthal-Kantor
triple system (abbreviated as $(\epsilon, \epsilon)$ BFKTS).  Let
$<x|y>$ to satisfy again Eq. (\ref{eq:onethree}), i.e.,

\begin{equation}
<x|y>\ = - \epsilon <y|x> \label{eq:threeone}
\end{equation}

\noindent for $\epsilon = \pm 1$. We write a triple product in $V$
now as a juxtaposition $xyz$. Suppose that it satisfies

\begin{subequations}\label{foo}
\begin{eqnarray}
&{\rm (i)}& \quad xyz - \epsilon zyx = 2<x|z>y \label{foo-a}
\label{eq:threetwoa}\\
\noalign{\vskip 4pt}%
&{\rm (ii)}& \quad xyz - \epsilon yxz = 2<x|y>z\label{foo-b}
\label{threetwob} \\
\noalign{\vskip 4pt}%
&{\rm (iii)}& \quad uv(xyz) = (uvx)yz + \epsilon x(vuy)z +
xy(uvz). \label{foo-c} \label{eq:threetwoc}
\end{eqnarray}
\end{subequations}

\noindent Then, any vector space $V$ with the triple product $xyz$
satisfy Eqs. (3.2) is called a $(\epsilon , \epsilon)$ BFKTS.

A simple example \cite{kamiya} is a triple product defined by

\begin{equation}
xyz = \ <x|z>y - \epsilon <x|y>z + \epsilon <y|z>x .
\label{eq:threethree}
\end{equation}

\noindent The reason why such a triple system is of interest is
due to the fact that we can construct Lie and anti-Lie triple
systems from them as follows.  We consider a larger vector space
$W$ by

\begin{equation}
W = V \oplus V . \label{eq:threefour}
\end{equation}

\noindent It is convenient to rewrite the generic element $w = x
\oplus y$ of $W$ as

\begin{equation}
w = \left( \begin{array}{c} x\\
\noalign{\vskip 6pt}%
y \end{array}\right), \quad x, \ y\ \in \ V , \label{eq:threefive}
\end{equation}

\noindent and introduce a new triple product in $W$ by

\begin{equation}
\left[ \left( \begin{array}{c} x_1 \\
\noalign{\vskip 6pt}%
x_2 \end{array} \right),
\left( \begin{array}{c} y_1\\
\noalign{\vskip 6pt}%
y_2 \end{array} \right),
\left( \begin{array}{c} z_1 \\
\noalign{\vskip 6pt}%
z_2 \end{array} \right) \right] : \  =
 \left( \begin{array}{c}
 w_1 \\
\noalign{\vskip 6pt}%
w_2 \end{array} \right) , \label{threesix}
\end{equation}

\noindent with

\begin{eqnarray}
w_1:\ &=& x_1 y_2z_1 - \epsilon y_1 x_2 z_1 + 2 \epsilon <x_1
|y_1>z_2 , \nonumber \\
\noalign{\vskip 4pt}%
w_2:\  &=& \epsilon y_2 x_1 z_2 - x_2 y_1 z_2 - 2 \epsilon <x_2
|y_2
>z_1 . \label{eq:threeseven}
\end{eqnarray}

\noindent Then, as a special case of a theorem \cite{kamiya} on a
more general Freudenthal-Kantor triple system, $W$ becomes a Lie
triple system for $\epsilon = +1$, and an anti-Lie triple system
for $\epsilon =-1$. Therefore, from any $(\epsilon , \epsilon)$
BFKTS, we can construct a Lie algebra for $\epsilon =+1$ and a Lie
superalgebra for $\epsilon =-1$ by the canonical construction
explained in section 2. Note that we have to let $\epsilon
\rightarrow - \epsilon$ in Eqs. (1.5) in order to now use the same
symbols for Lie and anti-Lie triple system.

Since we are interested in Lie superalgebras, we will consider
only the case of $(-1,-1)$ BFKTS hereafter.  Setting $\epsilon
=-1$ in Eq. (\ref{eq:threethree}), the resulting $(-1,-1)$ BFKTS
together with the canonical construction will give a Lie
superalgebra $osp(N,2)$ for $N = \dim V$. In order to obtain more
interesting Lie superalgebras, we will consider the following
examples in which the construction \cite{kamiya3} of exceptional
Lie superalgebras $D(2,1; \alpha)$, $G(3)$ and $F(4)$ are based.

\bigskip

\noindent \underbar{Example 1}

\bigskip

Let

\begin{equation}
V = \left\{ e_1 , e_2 , e_3, e_4 \right\} \label{eq:threeeight}
\end{equation}

\noindent with

\begin{equation}
<e_j |e_k>\ = \delta_{jk}, \quad (j,k = 1,2,3,4).
\label{eq:threenine}
\end{equation}

\noindent For an arbitrary constant $\sigma \ \in\ F$, a triple
product defined by

\begin{equation}
e_je_ke_\ell :\  = \sigma \sum^4_{m=1} \in_{jk \ell m} e_m -
\delta_{k \ell} e_j + \delta_{j \ell} e_k + \delta_{jk} e_\ell
\label{eq:threeten}
\end{equation}

\noindent gives a $(-1,-1)$ BFKTS. The resulting Lie superalgebra
is then $D(2,1 ; \alpha )$ with

\begin{equation}
\alpha = {1 - \sigma \over 1 + \sigma}. \label{eq:threeeleven}
\end{equation}

\bigskip

\noindent \underbar{Example 2}

\bigskip

Let $x \cdot y$ be an octonionic product in the octonion algebra
with

\begin{equation}
\overline x = 2 <e |x> e - x. \label{eq:threetwelve}
\end{equation}

\noindent Then, a triple product given by

\begin{equation}
xyz:\  = {1 \over 3} (x \cdot \overline y )\cdot z \ - {4 \over 3}
<y|z>x \ + {4 \over 3} <x|z>y\  - {2 \over 3} <x|y>z
\label{eq:threethirteen}
\end{equation}

\noindent defines a $(-1,-1)$ BFKTS. The corresponding Lie
superalgebra is the exceptional one $F(4)$ in the Kac's notation.

We can also construct an equivalent $(-1,-1)$ BFKTS in terms of
the 7-dimensional Dirac-Clifford algebra.

\begin{equation}
\gamma_\mu \gamma_\nu + \gamma_\nu \gamma_\mu = 2 \delta_{\mu \nu}
\quad (\mu , \nu = 1,2, \dots, 7). \label{eq:threefourteen}
\end{equation}

\noindent The multiplication table for the resulting $F(4)$
essentially gives the same one  given by Frappat et al.,
\cite{frap} as has been explained in reference 12.

\bigskip

\noindent \underbar{Example 3}

\bigskip

Let $x \cdot y$ again be the octonionic product but we restrict
ourselves  to a 7-dimensional sub-space

\begin{equation}
V = \{ x | x = \ {\rm octonion,\ with}\ <e|x>\ =0 \} ,
\label{eq:threefifteen}
\end{equation}

\noindent and set

$$ xyz:\  = - {1 \over 4} \{ (x \cdot y)\cdot z \ - \ x \cdot (y
\cdot z) \} \  - \ <y|z>x\  + \ <x|z>y \ + \ <x|y>z$$

\noindent which defines again a $(-1,-1)$ BFKTS. The resulting Lie
superalgebra is $G(3)$.

We can construct \cite{kam} also the strange Lie superalgebras
$P(n)$, and $Q(n)$ as well as the Cartan-type Lie superalgebras
$W(n)$ etc. from some other types of triple system. However we
will not go into detail.

\begin{theacknowledgments}
 This work is supported in part by the U.S. Department of Energy
 Grant DE-FG02-91ER40685.
\end{theacknowledgments}






\end{document}